\documentclass[sigconf]{acmart}

\AtBeginDocument{%
  \providecommand\BibTeX{{%
    \normalfont B\kern-0.5em{\scshape i\kern-0.25em b}\kern-0.8em\TeX}}}

\copyrightyear{2023}
\acmYear{2023}
\setcopyright{acmlicensed}\acmConference[]{To be published on the Proceedings of the International Conference on Evaluation and Assessment in Software Engineering}{June 14--16, 2023}{Oulu, Finland}
\acmBooktitle{Proceedings of the International Conference on Evaluation and Assessment in Software Engineering (EASE '23), June 14--16, 2023, Oulu, Finland}
\acmPrice{}
\acmDOI{}
\acmISBN{}

\acmConference[Submitted to EASE'23]{To be published on the International Symposium on Empirical Software Engineering and Measurement}{June 14--16, 2023}{Oulu, Finland}

\usepackage{array}
\usepackage{multirow}

\usepackage{xspace}

\newcommand{\ie}{\textit{i.e.}\xspace}
\newcommand{\eg}[0]{\textit{e.g.}\xspace}
\newcommand{\vs}[0]{\textit{vs.}\xspace}

\newcommand{\fig}{Fig.\xspace}

\begin{document}

\title{Measuring User Experience of Adaptive User Interfaces using EEG: A Replication Study}

\author{Daniel Gaspar-Figueiredo}
\affiliation{%
  \institution{ITI \& Universitat Politècnica de València}
  \city{Valencia}
  \country{Spain}
  \postcode{--}}
\email{dagasfi@epsa.upv.es}

\author{Silvia Abrahão}
\affiliation{%
  \institution{Universitat Politècnica de València}
  \city{Valencia}
  \country{Spain}}
\email{sabrahao@dsic.upv.es}

\author{Emilio Insfrán}
\affiliation{%
  \institution{Universitat Politècnica de València}
  \city{Valencia}
  \country{Spain}}
\email{einsfran@dsic.upv.es}

\author{Jean Vanderdonckt}
\affiliation{%
  \institution{Université catholique de Louvain}
  \city{Ottignies-Louvain-la-Neuve}
  \country{Belgium}}
\email{jean.vanderdonckt@uclouvain.be}

\renewcommand{\shortauthors}{Daniel Gaspar-Figueiredo et al.}

\begin{abstract}
  \textbf{Background}: Adaptive user interfaces 
  have the advantage of being able 
  to dynamically change their aspect and/or behaviour depending on the characteristics of the context of use,
  \ie
  to improve user experience. User experience is an important quality factor that has been primarily evaluated with classical measures (\eg effectiveness, efficiency, satisfaction), but
  to a lesser extent
  with physiological measures, such as emotion recognition, skin response, or brain activity.
  \textbf{Aim}: In a previous exploratory experiment involving users with different profiles and a wide range of ages, we analysed user experience in terms of cognitive load, engagement, attraction and memorisation when 
  employing
  twenty graphical adaptive menus through 
  the use of
  an Electroencephalogram (EEG) device. The results indicated that there were statistically significant differences for these four variables. 
  However, we considered that it was necessary to confirm or reject these findings using a more homogeneous group of users. 
  \textbf{Method}: We conducted an operational internal replication study with 40 participants. We also investigated the potential correlation between EEG signals and the participants' user experience ratings, such as their preferences.
  \textbf{Results}: The results of this experiment confirm that there are statistically significant differences between the EEG variables when the participants interact with the different adaptive menus. Moreover, there is a high correlation among the participants' user experience ratings and the EEG signals, and a trend regarding performance has emerged from our analysis.
  \textbf{Conclusions}: These findings suggest that EEG signals could be used to evaluate user experience. With regard to the menus studied, our results suggest that
  graphical menus with different structures and font types produce more differences in users' brain responses, while menus which use colours produce more similarities in users' brain responses. Several insights with which to improve users' experience of graphical adaptive menus are outlined.
\end{abstract}



\maketitle

\raggedbottom

\section{Introduction}

Software systems have become part of our daily lives, and the demand for more efficient and user-friendly systems has increased. Adaptive systems are designed in order to adjust to the changing needs and preferences of their users.
User Interface (UI) adaptation is, in particular, a crucial aspect of adaptive systems as it allows the system to adjust its interface in order to better suit the user's needs. Some users may need or prefer different 
features of the UI (\eg layout, fonts),
and by adapting to these preferences, the system can provide a more efficient and satisfying user experience (UX). 
Aspects of user experience can be measured by analysing physiological measures that represent different human affective states (\eg pleasure, stress, relaxation) when performing a particular task. 
This kind of measures has received increasing attention from the Software Engineering (SE) community in the last few years. 
 A recent systematic literature review on the use of physiological measures in SE~\cite{weber2021brainSLR} revealed that several empirical studies have been conducted in order to measure the cognitive load in SE activities, such as: code comprehension, code inspection, programming, and bug fixing.
 However, only a few have reported experiments with which to capture end-user emotions and feelings. In addition to new experiments, it is also necessary to carry out replications that will increase the body of knowledge concerning the usefulness of these physiological measures as regards supporting SE tasks.

    In a previous study~\cite{prevExperiment}, we performed an exploratory experiment in order to assess the impact of adaptive user interfaces on the cognitive load and user experience, both of which were measured using EEG signals. Since designing experiments that study the effect of a wide range of UI elements on the UX may imply risks, the goal of this experiment was to assess whether there are significant differences among the physiological responses of a group of 40 participants when using twenty different types of graphical adaptive menus selected from the catalogue reported in~\cite{vanderdonckt2019Menus}. The results of the experiment showed the feasibility of measuring cognitive load and user experience using EEG. The factorial analysis showed that there were significant differences between the graphical menus in terms of cognitive load, engagement, attraction and memorisation. Nevertheless, the participants' background was not homogeneous, since there were different profiles and a wide range of ages.  
    
    In this paper, we present an operational internal replication of this experiment with a group of 40 Computer Science students who were enrolled on a Master’s degree programme at the 
    \textit{Universitat Politècnica de València (UPV)}
    The goal of this replication was to verify the findings of the baseline experiment in a different context and to expand the research by adding a new research question related to the correlation between EEG signals and the participants' user experience ratings.
    
    This paper is structured as follows. Section 2 and 3 describe the background and related work, respectively, while Section 4 describes the replication study. Section 5 describes the results obtained and Section 6 presents the conclusions.


\section{Background}

    \begin{figure*}
        \centering
        \includegraphics[width=0.982\linewidth]{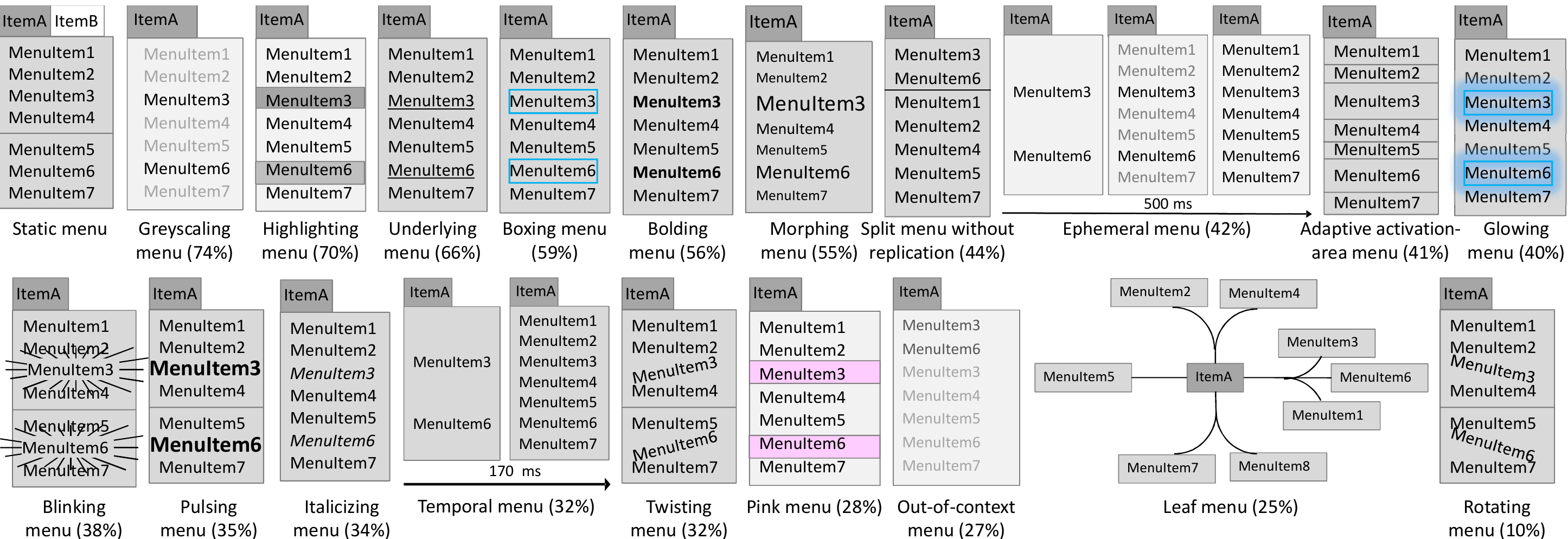}
        \caption{The twenty 
        menus used in the experiment. The percentage shows the preference ratings as reported in~\cite{vanderdonckt2019Menus}.
        }
        \label{fig:usedMenus}
    \end{figure*}

\subsection{Adaptive User Interfaces}
\label{ref:AUI}

    User needs, wishes, and preferences may change over time, and software systems should be able to identify those circumstances and adapt themselves to meet the new needs or preferences. Adaptive systems are systems that can modify aspects of their structure or functionalities in order to satisfy different users' needs and track their changes over time. 
    Adaptive user interfaces are adaptive systems that are capable of adjusting the display, thus adapting the organisation or presentation of the User Interface (UI) functionality in response to certain characteristics of the user's context.
    
    Various strategies with which to support the adaptation of UIs have been proposed in literature. These strategies cover the UI as a whole, or certain elements of the user interface. In a recent study~\cite{vanderdonckt2019Menus}, a set of 49 graphical adaptive menus was collected which provided a new exploration of a UI design space on the basis of Bertin's eight visual variables (\ie position, size, shape, value, colour, orientation, texture, and motion)~\cite{bertin1967semiologie}. This design space allowed the authors to propose a set of graphical adaptive menus that can be expanded and compared. Graphical adaptive menus are graphical user interface menus whose predicted items of immediate use can be automatically rendered in a prediction window (via a subset of menu items resulting from a prediction scheme). The different graphical adaptive menus highlight the prediction window with the purpose of improving users' efficiency when using menus. However, experiments with which to assess how different design choices affect the user experience are required.

    In this paper, we present a replication of an experiment that compares a sub-set of twenty graphical adaptive menus proposed in \cite{vanderdonckt2019Menus}. The selected menus cover most of Bertin's visual variables~\cite{bertin1967semiologie}. We specifically selected Adaptive activation area, Blinking, Bolding, Boxing, Ephemeral, Glowing, Greyscale, Highlighting, Italicising, Leaf, Morphing, Out-of-context disappearing, Pink, Pulsing, Rotating, Split without replication, Temporal, Twisting, and Underlying. As a baseline, we also used the Static menu, which does not use the prediction window and does not highlight any specific menu item.
    Figure \ref{fig:usedMenus} shows all the menus selected.
    For example, the Underlying menu is a value-changing menu (\ie menus that change the font of the prediction window, underlining the menu items in order to highlight them),
    while a Rotating menu is a motion-changing menu (\ie menus that move menu items in order to highlight them) and Leaf menu is a menu that contains unusual shapes.
    



\subsection{User Experience}

     The ISO 9241-210~\cite{ISO9241} defines the term user experience (UX) as \textit{"a person’s perceptions and responses that result from the use and/or anticipated use of a product, system, or service"}. User experience, therefore, includes all the effects the use of an interface may have on the user, before, during, and after use.
    The two major categories of UX evaluation methods are termed as: \textit{subjective}, which are usually measured through interviews, surveys, ratings and questionnaires, 
    and \textit{objective}, which are usually measured by employing a user performance measure, a physiological and neurological attribute response, etc.~\cite{zaki2021SLR}. 
    However, the subjective approaches have limitations that may provide biased information. Moreover, the responses can be influenced by how users remember their experience and not by the real experience itself~\cite{kujala2011ux}.
    This inaccuracy in UX measurement has led us to seek alternatives. It is already known that physiological reactions such as blinking, sweating, eye movement, or brain activity can be used to objectively measure UX \cite{bergstrom2014physiological}. 
    
      
    In this replication, we used both EEG and post-experimental questionnaires to measure the UX produced when using the twenty adaptive menus mentioned in Section~\ref{ref:AUI}. We decided to use post-experimental questionnaires, despite the disadvantages mentioned above, because we also wished to study whether these subjective measures of UX correlated with the objective measures.

\subsection{Brain Activity Measurement using EEG}
\label{ref:UxEEG}

    
    An electroencephalogram (EEG) is a technique that measures brain activity by recording changes in electrical activity, also called brainwaves, generated by neurons via electrodes placed on the scalp. 
    %
    Brainwaves can be captured and analysed to obtain the user's emotional status while doing any task. This means that it is possible to can detect when a user is concentrated, relaxed, excited, etc.
    
     EEG data can be captured by using a cap or a headset with electrodes. Several factors make the devices different, such as the type of electrode, the size of the cap or the type of device. These devices are usually connected to an amplifier, which may also vary (\eg sample rating, bandwidth, resolution). 
     The brainwaves captured are characterised on the basis of multiple factors. The characteristics most frequently used are the frequency and the location of the electrodes. The frequency is typically divided in 5 groups: delta (0.5 to 4Hz), theta (4 to 7Hz), alpha (8 to 12Hz), sigma (12 to 16Hz) and beta (13 to 30Hz). 
    Furthermore, the analysis of different brain zones has been found to provide information on users' emotions. For example, several studies with which to infer the role played by frontal brain activity in emotion have been conducted~\cite{Allen2004Issues}, and differences between left and right frontal cortical activity are similarly associated with positive emotions like enthusiasm or negative emotions like anger~\cite{HARMONJONES2010role}.
    Although brain zones are important in the analysis of emotions, brain frequencies are equally important.
    Alpha and theta waves have been found to be correlated with cognitive and memory performance \cite{Klimesch1999EEG}, and if beta bandwidth is also considered, it is possible to measure user engagement \cite{Mikulka2002Effects}.
       
    Finally, each person may have different brain responses in relaxing or stressful situations. A baseline sample is, therefore, necessary for each participant in order to establish their relaxed and stressed states. This baseline sample is called a \textit{calibration sample}. For example, the most frequently used calibration sample for a relaxed state is that in which the participants are asked to close their eyes in order to measure their brain activity without physiological impulses.

\section{Related work}








EEG has, over the last few years, been used in educational contexts to measure cognitive load. For example, in \cite{molleri2019experiences} the authors measured the cognitive load produced by the review of proposals for Master's degree theses. They used an EEG headset to measure the participants' attention rate while reviewing the proposals. The main goal of this work was to provide guidelines on how to plan and execute this type of experiments. Similarly, in \cite{antonenko2010eegCL}, EEG was also used to assess the cognitive load produced in the context of hypertext and multimedia learning. The main goal was to provide evidence for the feasibility of using EEG in educational research in order to collect and analyse the cognitive load to test the effectiveness and improve the design of learning materials. In both studies, the authors concluded that EEG is a good way in which to 
objectively measure the cognitive load produced by different tasks.

Moreover, some studies have explored the use of psychometric data to measure cognitive load in SE tasks such as code comprehension, code readability, and error detection. For example, \citet{siegmund2014understanding} used fMRI (functional Magnetic Resonance Imaging) to determine the cognitive load registered by code comprehension tasks and locate syntax errors with which to contrast them. They found a different activation pattern in five regions of the brain related to working memory, attention, and language processing.
In another study, \citet{fakhoury2018effect} used fNIRS (functional Near-Infrared Spectroscopy) combined with eye-tracking to explore the effect of poor source code lexicon and readability on developers' cognitive load. The authors concluded that the presence of linguistic anti-patterns in source code significantly increases the developers' cognitive load. 
Similarly, \citet{Peitek:2018} found some limitations when using only fMRI, and consequently also decided to join fMRI data to eye tracking data in order to attain a better understanding of the cognitive processes during programme comprehension.



Moreover, as acknowledged by \citet{feldt2008towards} empirical SE studies should focus not only on the developer's perception of the system, but also on that of the end user. The user experience is usually measured by means of interviews or questionnaires after the system has been used. 
For example, cognitive load was traditionally measured using the NASA-TLX questionnaire to correlate it with user satisfaction~\cite{schmutz2009cognitive}. However, these responses may be subject to inaccurate user recall or other subjective factors, leading to imprecise measurements~\cite{kujala2011ux}.


Some experiments are starting to include 
physiological data in their data collection. For example, \cite{lee2009brain} used fMRI and EEG to describe how the brain reacts when users see different designs. The authors concluded that fMRI showed that the perception of different feelings towards designs is associated with the frontal and occipital lobes, and the EEG showed that the human brain responds sooner and stronger as regards its perception of bad feelings.


In another study 
\cite{hou2017eeg} the authors evaluated an Air Traffic Control system using EEG-based tools to monitor and record the brain states of air traffic controllers. 
The authors also analysed the relationship between the cognitive load calculated using the traditional NASA-TLX method and that calculated using the method with which to label EEG data. They found that the data obtained from most of the simulations the data were highly correlated. 
Some studies that compare adaptive user interfaces have also been performed. For example, Gajos et al.~\cite{gajos2006exploring} implemented and evaluated three graphical adaptive 
UIs
in two experiments together with a non-adaptive baseline. The authors concluded that adaptive UIs have an impact on users depending on their particular properties. Those interfaces that frequently  duplicate (rather than moving) tend to improve users' performance and satisfaction. Other researchers \cite{findlater2009design}, when focusing on adaptive graphical menus, compared two adaptive interface designs and found that these provided more consistently positive results in terms of performance and user satisfaction. 

The experiments mentioned above focus on detecting differences between brainwave frequencies and the cognitive load. As mentioned in \cite{hou2017eeg}, emotions can also change, and this is also important in UX evaluations. In this study, data from EEG and questionnaires are used to measure not only the cognitive load, but also the attraction, the memorisation effort, and the user engagement produced by the interaction with different user interfaces.

\section{Experimentation}

    We first present an overview of the baseline experiment, followed by a description of the design and execution of the internal replication.

\subsection{Overview of the baseline experiment}

    Different UIs may lead to different emotions and mental effort, and, overall, to each user having a different UX. There are several types of UI elements that could be considered. In order to focus on one type of element, we considered the different types of graphical menus proposed in  \cite{vanderdonckt2019Menus} (see Section~\ref{ref:AUI}).
    This set of graphical adaptive menus has been subjected to a preference analysis, but no empirical study has been performed in order to compare the impact of the different menu types on UX. We accordingly performed an exploratory experiment with which to analyse this impact \cite{prevExperiment}, and our findings are briefly introduced in the following sections.

\subsubsection{Context, goal and design of the experiment}

    The baseline experiment was conducted with 40 volunteers (20 males and 20 females) with diverse backgrounds (\eg software engineers, mechanical engineers, financial experts, healthcare professionals) and a wide range of ages (from 18 to 64 years old, $M{=}39.55$, $SD{=}15.26$, $Mdn{=}41$). The volunteers were recruited by means of an open call for participation at the 
    \textit{UPV}
    using a snowball procedure.

    The goal of the experiment was to 
        \textbf {analyse} a set of adaptive graphical menus \textbf{for the purpose of} comparing them with respect to the user experience produced in terms of cognitive load, engagement, memorisation and attraction \textbf{from the point of view of} both researchers and software designers \textbf{in the context of} members of an end-user group at the 
        \textit{Universitat Politècnica de València}
        and their contacts.
    To this end, we investigated the effect of using menus with different properties (\eg position-changing, orientation-changing or size-changing).
    We selected a sub-set of twenty menus covering most of the properties. We used an EEG headset to gather the UX measures from the brain activity. In particular, we obtained four variables related to UX: cognitive load, engagement, memorisation and attraction. 
    These UX measures were obtained by registering the participants' brainwaves while they were using twenty different menus. The completion time was also registered. The objective of the alternative hypotheses tested in the experiment was to prove whether there were differences among the metrics obtained for the menus used.

    The design of the experiment was a within-subjects design, in which each participant used all the menus. Each participant used a total of 20 menus in a random order.
    Moreover, the menus were randomly assigned to two different contexts: a mail manager menu and a web browser settings menu. The menu item to be selected (experimental task) was also random. 

\subsubsection{Experiment Results}

The results of the experiment showed that using different types of graphical adaptive menus can produce statistically significant differences in the UX as regards the cognitive load, engagement, memorisation, attraction and completion time (\ie cognitive load p-value=0.000, engagement p-value=0.000, memorisation p-value=0.000, attraction p-value=0.000, completion time p-value=0.000). Dynamic Time Warping (DTW) was used in this experiment in order to compare the time series registered for each EEG metric. 
 These findings suggest that menus with atypical structures, different font styles, movement, and increased area of menu items can have a more subjective and user-dependent effect on the UX, while menus with temporality or colour tend to be more objective and easier to use and learn. Suggesting that some menus may be more suitable for certain tasks or contexts than others.
For more details about the baseline experiment please refer to \cite{prevExperiment}. 

\subsection{Internal replication}

We carried out an operational internal replication of the baseline experiment~\cite{prevExperiment}. The same experimental protocol was applied with the same operationalization and experimenters, but to a different population~\cite{GOMEZ:2014}. We also added a new research question, hypotheses and variables in order to attain more insights into the results. The purpose of this replication was to test the extent to which the study results could be generalised to other populations (\ie a more homogeneous group of users). Replications are needed in order to increase confidence in the validity of the experiment’s conclusion. An additional motivation is that the sample size of the baseline experiment can affect the magnitude of the effect (in this case, the strength of the relationship between the type of menu and the UX measures of that population). It is, therefore, convenient to run internal replications so as to increase the sample size, thus improving the effectiveness of confirming the experiment hypotheses.

\subsubsection{Research questions}

    The goal of this study was to determine how graphical adaptive menus impact on user experience, measured using an EEG device. Furthermore, we wished to assess whether the UX measures gathered using EEG were correlated with the cognitive data obtained through the use of traditional UX questionnaires. The perspective was that of software designers interested in obtaining empirical evidence about the effect of different graphical adaptive menus on the emotions and cognitive load of end users. The objective of the study was, therefore, to answer the following research questions:
    
    \begin{itemize}
        \item RQ$_{1}$: Do the 20 graphical adaptive menus have a different influence on user experience?
        \item RQ$_{2}$: Does user experience measured using EEG signals correlate with the subjective ratings obtained using traditional questionnaires?
    \end{itemize}
    
\subsubsection{EEG device, variables and questionnaires selected}
\label{sec:BCI,VAR}
    In section \ref{ref:AUI}, we presented the menus selected in the baseline experiment (see \fig \ref{fig:usedMenus}). 
    These menus mostly cover all of the properties in the design space (\eg position, motion or colour).
    These menus have different properties, and we expected that this would produce differences among the multiple participants' user experience. For this empirical experiment, there was one independent variable with 20 possible values.
    The dependent variables can be divided in three types: emotion-based, perception-based and performance-based variables. 
    Emotion-based variables assess participants' emotions and feelings objectively. In Section~\ref{ref:UxEEG}, we presented how brain activity can be measured in order to obtain different relevant aspects of UX. In this experiment, the emotion-based dependent variables employed were the following, as defined in~\cite{bitbrain:b2020}:
    
  \begin{itemize}
        \item \textbf{Cognitive load}: measures the participants' concentration when presented with stimuli or during experiences. 
        It is expressed as a percentage. Values close to 0\% indicate that the participants are very distracted, while values close to 100\% indicate that they are very attentive to the stimulus.
        
        \item \textbf{Engagement:} measures the degree of involvement or connection between the participant and the stimulus or task. 
        It is expressed as a percentage. A value close to 0\% indicates that there is no connection to the stimuli while a value close to 100\% indicates high engagement with the stimuli or task.

        \item \textbf{Attraction}: measures the degree of attraction experienced in response to a stimulus, 
        from a positive (pleasant) reaction to a negative (unpleasant) reaction. It is expressed as a percentage. A value close to 100\% positive (pleasant) or 100\% negative (unpleasant) is equivalent to the response measured as baseline during the calibration phase.
        
        \item \textbf{Memorisation}: measures the intensity of cognitive processes related to the formation of future memories during the presentation of stimuli.
        Captures the degree of memory storage, encoding, and retention. It is expressed as a percentage. A value of 0\% indicates that the probability that the stimulus will be remembered is low, while a value close to 100\% indicates a high possibility that the stimulus will be retained in the participants' memory.

    \end{itemize}
    
    Perception-based variables subjectively assess participants’ perceptions of their emotions.
    In this experiment, the perception-based dependent variables were specifically the \textbf{Perceived Cognitive Load (PCL)} and the \textbf{Perceived Attraction (PA)}.
    
    Performance-based variables assess participants' performance when selecting and interacting with different menus. The specific performance-based variable  employed in this experiment was the \textbf{Completion time (CT)}. This measure reflects the amount of time it takes for a user to complete a task.
    It is a measure of efficiency and can have a significant impact on the overall user experience. 

\begin{figure}
  \centering
  \includegraphics[width=\linewidth]{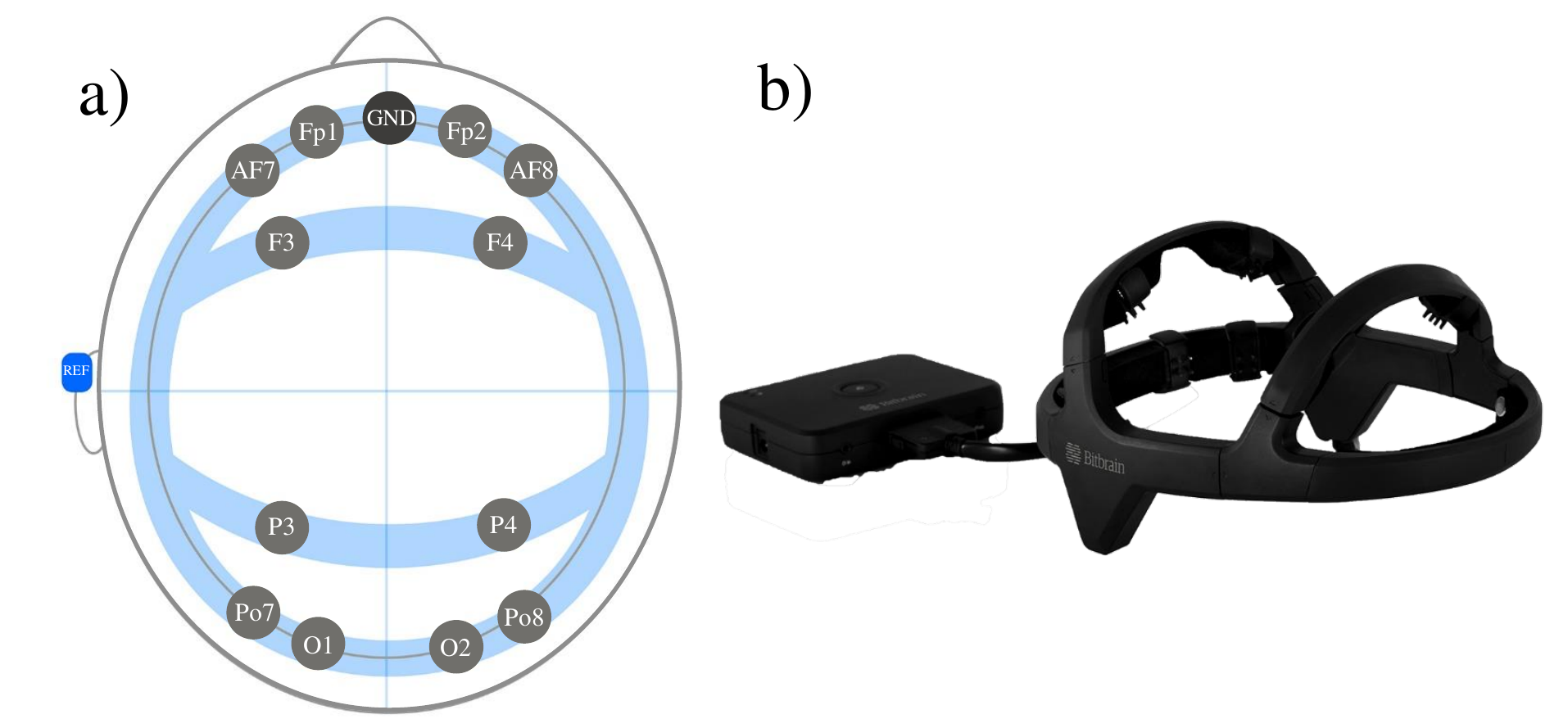}
  
  \caption{a): The location map of the 12 electrodes based on international 10-10 system. b): The Diadem BCI.}
  \label{fig:bci}
\end{figure}

    In order to obtain the emotion-based dependent variable, Bitbrain's Diadem headset~\footnote{https://www.bitbrain.com/neurotechnology-products/dry-eeg/diadem}
    was used. This is a portable non-intrusive EEG headset that is widely used for neuroscience research and brain computer interfaces (BCI). This device has twelve dry electrodes placed at positions AF7, Fp1, Fp2, AF8, F3, F4, P3, P4, PO7, O1, O2 and PO8 according to the international 10-10 standard \cite{electrode1994guideline}, as shown in Figure \ref{fig:bci}. This headset also uses the left earlobe (A1, REF) as a reference electrode and the centre of the head (Fpz, GND) as the ground. It is optimised to estimate emotional and cognitive states (\ie it measures brain activity in the prefrontal, frontal, parietal, and occipital brain areas). 
    The headset is wired to an amplifier that communicates with the researchers' computer via Bluetooth. The transition from the brainwaves to UX measures, was computed using the Bitbrain Sennscloud processing service~\cite{bitbrain:b2020}. This service uses calibration data and brain activity data in order to compute these four metrics. They compute these metrics on the basis of prior research into EEG data analysis, as seen in Section \ref{ref:UxEEG}.
    This is, for example, done for Cognitive load \cite{Klimesch1999EEG}, Engagement \cite{Mikulka2002Effects}, 
    Memorisation \cite{Long2014Subsequent} and
    Attraction (\cite{Allen2004Issues}, \cite{HARMONJONES2010role}).



    Furthermore, in order to obtain the perception-based variables, the NASA-Task Load Index (NASA-TLX) and the User Experience Questionnaire - Short (UEQ-S) were used. These questionnaires are standardised and are both widely applied and empirically validated survey questionnaires. 
    The NASA-TLX questionnaire is a tool with which to measure subjective mental cognitive load. It rates performance across six dimensions (mental demand, physical demand, temporal demand, effort, performance, and frustration) so as to determine an overall cognitive load rating. This questionnaire is divided into two parts. First, each dimension is rated within a range of 100 points with five-point steps. The objective of the second part is to create an individual weighting of these dimensions by allowing the participants to compare them in a pairwise manner on the basis of their perceived importance (\ie the participants choose which dimension is most relevant to cognitive load). 
    The UEQ-S is used to obtain the hedonic quality (HQ), which is the best predictor of the attraction of an experience \cite{tuch2016leisure}, and the pragmatic quality (PQ), which is also relevant for UX measurements. It consists of eight items that are grouped into two scales (\ie hedonic and pragmatic quality). Each item has a positive and a negative value that must be rated from 1 (which is the negative value of the item) to 7 (which is the positive value of the item), with 4 being the neutral value. According to the UEQ-S handbook, the items 1 to 4 are related to PQ and items 5 to 8 are related to HQ.
    The items used and their values, along with the questionnaires used in this experiment, can be found at: 
    https://issi-dsic.github.io/


    Finally, the performance-based variable was obtained by using, the recordings of the sessions from the experiment. These recordings were then analysed in order to determine the amount of time it took for the participants to complete the tasks assigned to them (\ie completion time) while using the different menu types.

\subsubsection{Hypothesis formulation}
\label{ref:hypothesisForm}
    The following null hypotheses, which are related to the emotion-based variables, were defined in order to verify the first research question, namely whether the use of twenty graphical adaptive menus has a different impact on UX:
    
\begin{itemize}
    \item $H_{n11}$: There are no significant differences in the users' \textbf{cognitive load} when using different adaptive menus.
    
    \item $H_{n12}$: There are no significant differences in the users' \textbf{engagement} when using different adaptive menus.
    
    \item $H_{n13}$: There are no significant differences in the users' \textbf{attraction} when using different adaptive menus.
    
    \item $H_{n14}$: There are no significant differences in the users' \textbf{memorisation} when using different adaptive menus.
      
\end{itemize} 

    The following hypothesis was also formulated for the first research question, but concerned the performance-based variable:
    
\begin{itemize}
    \item $H_{n15}$: There are no significant differences among the users' \textbf{completion times} when using different adaptive menus.
\end{itemize}

    The following null hypotheses were similarly defined in order to verify the second research question, namely whether user experience measured using EEG signals correlates with the subjective ratings obtained using traditional questionnaires:
    
\begin{itemize}
    \item $H_{n21}$: There is no correlation between the \textbf{cognitive load} and the \textbf{perceived cognitive load}. 
    
    \item $H_{n22}$: There is no correlation between the \textbf{attraction} and the \textbf{perceived attraction}.
    
\end{itemize}

    The goal of the statistical analysis was to reject these hypotheses and possibly to accept the alternative ones (\eg $H_{a11}$ = $\neg H_{n11}$ ). All the hypotheses are two-sided because we did not postulate that any effect would occur as a result of graphical adaptive menu usage.

   Although we could have obtained four UX measures using the EEG device selected, and one measure to measure the performance, we nevertheless decided to correlate cognitive load and attraction only, as these are the most important cognitive and emotional states when measuring and modelling UX~\cite{tuch2016leisure}.
   
   Moreover, in order to mitigate a potential maturation threat (\ie subject tiredness) we limited the number of questionnaires that the participants had to complete. We asked them to fill in the NASA-TLX and UEQ-S for just six of the graphical adaptive menus. They specifically filled in the questionnaire regarding 3 menus that were more similar to those used in the baseline experiment, and the 3 that were more different (\ie Blinking, Split without replication, Out-context disappearing and Morphing, Italicising and Leaf).
   

    
    
    

\subsubsection{Context and subject selection}

    The participants included in this study were selected by means of convenience sampling. They were 40 students enrolled on a Master’s degree in Computer Science at the 
    \textit{Universitat Politècnica de València}
    Specifically, 35 of them were male and 5 were female,
    aged from 20 to 40 (mean of 24). 
    They all had experience of using UIs in their studies, professional occupations or hobbies.
    The number of participants is in line with previous UX studies that have used EEG as a measurement device. It is not, as yet, possible to determine a sample size for UX tests performed with physiological monitoring, but it has been shown that the median sample size in EEG studies is 18 participants~\cite{ApraizIriarte2021Evaluating}.
    All the participants were volunteers and were made aware of the practical and pedagogical purposes of the experiment, but the research questions were not disclosed to them. The participants were not rewarded for their effort.

\subsubsection{Design of the replication}

 Our experimental design was a within-subjects design, in which each participant used all the graphical adaptive menus. We carried out one EEG session per participant. We represented the different graphical adaptive menus by using an interactive slide presentation. Our menu representations were based on two different domains: a mail manager software and a web browser. Each participant was asked to use all types of menus and in a random order so as to ensure that the results of a menu were not influenced by the order in which they were used. We specifically prepared an interactive slide presentation for each participant in which the twenty menu types and the two domains were randomised (\ie 10 menu types with each domain).
 In order to answer the $RQ1$ we used the information regarding the interaction with the 20 menus, while $RQ2$ was answered by using only information concerning the interaction and questionnaire responses for 6 of the 20 menus, as explained in Section \ref{ref:hypothesisForm}.

\subsubsection{Preparation and Data analysis}
\label{ref:prepDA}

Prior to the experiment, each participant signed a consent form and a demographics form. They were then given a set of instructions to inform them about the experiment protocol. An experimenter was also present to answer any questions. After the forms had been filled in, the session continued with the headset placement and calibration phase that lasted about eight minutes. It is mandatory to establish a baseline brain activity for each participant in order to compute the UX measures. The calibration phase was, therefore, composed of five steps:

\begin{enumerate}
    \item Calibration task familiarisation (1 min.): The participant learns to perform the calibration task. 
    
    \item Closed eyes baseline (2 min.): The goal is to decouple the emotional response from one block to the next and to establish a null baseline for EEG and biosensing data. This is a break during which the participants clear their emotional responses. The participants are asked to close their eyes.
    
    \item Open eyes baseline (2 min.): The goal is to establish a baseline measurement in EEG and biosensing. The participants are instructed to enter a rest state with their eyes open.
    
    \item Calibration baseline (1 min.): The goal is to determine the range of physiological responses for the individual, and normalise across all participants. 
     

    
    \item Closed eyes washout (2 min.): The goal is to decouple the emotional response from the calibration to the experiment. 
    
\end{enumerate}

In order to familiarise the participants with the menu types, a practice trial with a different graphical adaptive menu was performed as many times as necessary in a different context. Once familiarised, and after the resting phase, they were able to start the experimental task: selecting a menu item for each one of the twenty graphical adaptive menus. The experimental task is composed mainly of three repeated steps: 1) Use the graphical adaptive menu with no time limit; 2) Resting phase (ten seconds) and 3) Open eyes. In step 1, one random graphical adaptive menu (out of the 20 menus) was randomly shown on the screen. The menu was first hidden, and there participants were asked to select a specific menu item from the predicted window that would be highlighted by the menu. After expanding the menu, all the menu options were then displayed, and the participants had to select the menu item requested previously. In step 2, after selecting the menu item, the participants were asked to close their eyes and do nothing for ten seconds. This resting phase was used to wash out emotions (e.g. to separate the emotions felt towards one menu from those felt towards another). In step 3, the participants were asked to open their eyes again, and repeat the process from step 1. All mouse input and brainwaves were recorded throughout the process. Finally, the participants filled in the two questionnaires with which to assess their perceived cognitive load and perceived attraction for six out of the twenty menu types. As explained in Section~\ref{ref:hypothesisForm}, we selected the six menus that were found to have the most/least differences according to the baseline experiment~\cite{prevExperiment}.

The results of the experiment were collected using the Bitbrain SennsLab software and the questionnaires. SennsLab records the EEG signals in real-time (i.e. each participant’s brainwaves), and they are then processed using SennsCloud to compute the UX measures in relation to the brain zones, brainwave bandwidth, and calibration samples obtained from the EEG data. We then used Python, Excel and R Studio to analyse the data collected. Raw brainwave data was subsequently processed in order to obtain the four emotion-based variables in the form of time series data. These were compared by using the Dynamic Time Warping (DTW) distance measure, a method that allows the comparison of non-uniform time intervals and can handle time series of different lengths~\cite{Muller:2007}.
We computed DTW distances in order to assess similarity or differences among the participants’ emotions while using the menus; small distances suggest similarity while large distances suggest differences.
DTW distance is not constrained to the time series scale.
We obtained the mean distance for each participant and menu by: 1) Grouping time series data by menu in order to compare the participants; 2) Computing the distance matrix for each menu and variable, and 3) Calculating the mean distance per participant by taking the mean of each matrix column. This process was repeated for each UX measure and menu. The results obtained showed the differences and similarities among the participants’ brain responses.
For more details on this process, please refer to~\cite{prevExperiment}.

We then used descriptive statistics, histogram plots, and statistical tests to analyse the data collected. As is usual, in all the tests, we accepted a probability of 5\% of committing a Type-I Error, \ie rejecting the null hypothesis when it is in fact true. The data analysis was carried out as follows:

\begin{enumerate}
  \item We first carried out a descriptive study of the measures for the dependent variables.
  
  \item We analysed the characteristics of the data in order to determine which test would be most appropriate to test our hypotheses. Since the sample size was less than 50, we applied the Shapiro–Wilk test to test the normality of the data, and the Brown-Forsythe Levene-type test to determine the homogeneity of any variances.
  
  \item The results of the tests were then employed as a basis on which to test the null hypotheses formulated. When the data were normally distributed and the variances were homogeneous, we used one-way Analysis of Variance (ANOVA) to analyse the data by considering the menu type as a main factor.
  When the ANOVA assumptions could not be satisfied, we used the Kruskal–Wallis test) to compare the mean averages of the twenty treatments. 
  
  \item We analysed both the NASA-TLX and the UEQ-S questionnaires. With the NASA-TLX, we used an Excel file which computes the subjective cognitive load on the basis of 
  users' responses. The UEQ-S data was analysed using the UEQ Data Analysis Tool, which is available 
  on the UEQ homepage.
  This tool is also an Excel file which facilitates data analysis. We first entered the data from the UEQ-S onto the Data worksheet and the tool then calculated all the statistics required in order to interpret the results. 

  \item We analysed the correlation of the UX measures obtained using the EEG signals with the perceived UX measures collected using the questionnaires.
  We used Pearson’s correlation coefficient to analyse the data by comparing the means of the objective measure with the corresponding subjective measure (\ie Attraction and Perceived Attraction). When the normality assumption did not hold, we used Spearman’s correlation coefficient. 

  \item The statistical significance of the experiment was complemented with the magnitude of its effects. Cliff’s $\delta$ estimates~\cite{Cliff1993Dominance} 
  were, therefore, obtained, with a confidence interval of 95\%. These measures are recommended when dealing with ordinal scale data \cite{Kitchenham2017robust}. Moreover, the non-parametric nature of Cliff’s $\delta$ estimates serves to reduce the influence of distribution shape, differences in dispersion, and extreme values. The magnitude of the effect was assessed using the thresholds
  |d| < 0.112 “negligible ”, |d| < 0.276 "small"  |d| < 0.428 "medium", otherwise "large".
  
\end{enumerate}

\section{Results}

\begin{figure}
  \centering
  \includegraphics[width=0.79\linewidth]{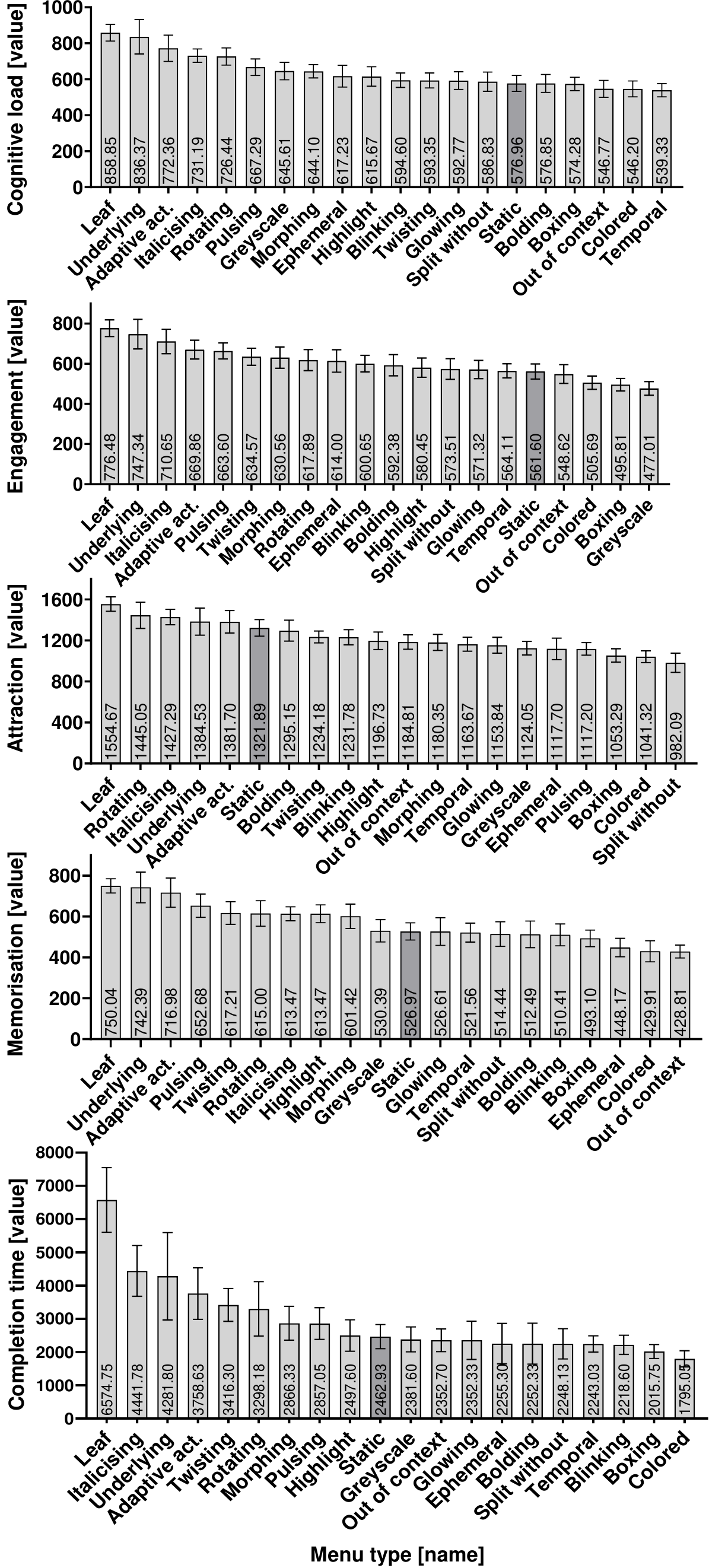}
  \caption{Measurements obtained for each menu type, in decreasing order of their mean value, in comparison to the static menu (darker colour) as a baseline.}
  \label{fig:emotionBasedResults1}
\end{figure}

  The descriptive analysis showed that participants' the emotion-based and performance-based variables varied across the different graphical adaptive menus. Figure \ref{fig:emotionBasedResults1} visually presents the results obtained for each menu type. The error bars show 95\% confidence intervals. We used the \textit{Static} menu as a baseline because its properties are constant. This figure shows the DTW distance measured for each of the emotion-based variables. These measures are not constrained to the scale of the variables.
  


    At a glance, it will be noted that menus with atypical structures (\textit{Leaf}), with different font styles (\textit{Italicising} and \textit{Underlying}) and menus with movement (\textit{Rotating}) could influence very differently, in general, among users. 
    In contrast, the registered metrics were more similar between participants when using menus that use temporality (\textit{Temporal} and \textit{Out of context disappearing}), colour-changing properties (\textit{Coloured}, \textit{Boxing} and \textit{Greyscale}). 
    Note that, with regard to attraction, if no properties are changed (\textit{Static} menu) the participants reaction was very different. 
    This could indicate that the participants had different preferences from a starting point and varying the properties produced more similarities among them.

\subsection{RQ1: Influence of adaptive menus}

    The Shapiro-Wilk test revealed that none of the emotion-based and performance-based variables followed a normal distribution (\ie p-value < 0.05). The Brown-Forsythe Levene-type test revealed that the emotion-based variables were homogeneous for all the menu types, and that the completion time was not homogeneous for any of the menus. 
    Table \ref{tab:hypothesis-test} shows the statistical tests that were applied to each emotion-based and completion time variable and their respective significance level. The Kruskal-Wallis test revealed that the null hypotheses $H_{n11}$, $H_{n12}$, $H_{n13}$, $H_{n14}$ and $H_{n15}$ could be rejected (\ie cognitive load p-value = 0.000, engagement p-value = 0.000, attraction p-value = 0.000, memorisation p-value = 0.000, completion time p-value = 0.000). These results indicate that there are statistically significant differences among the participants' cognitive load, engagement, attraction, memorisation, and completion time when using the twenty graphical adaptive menus.
    
    We additionally found practical significance for these hypotheses by using Cliff’s $\delta$. As shown in Table \ref{tab:hypothesis-test}, cognitive load, memorisation and the completion time had a large effect, size while attraction and engagement had a medium effect size. 
    
    

\subsection{RQ2: Correlation analysis}




\begin{figure*}
  \centering
  \includegraphics[width=0.98\linewidth]{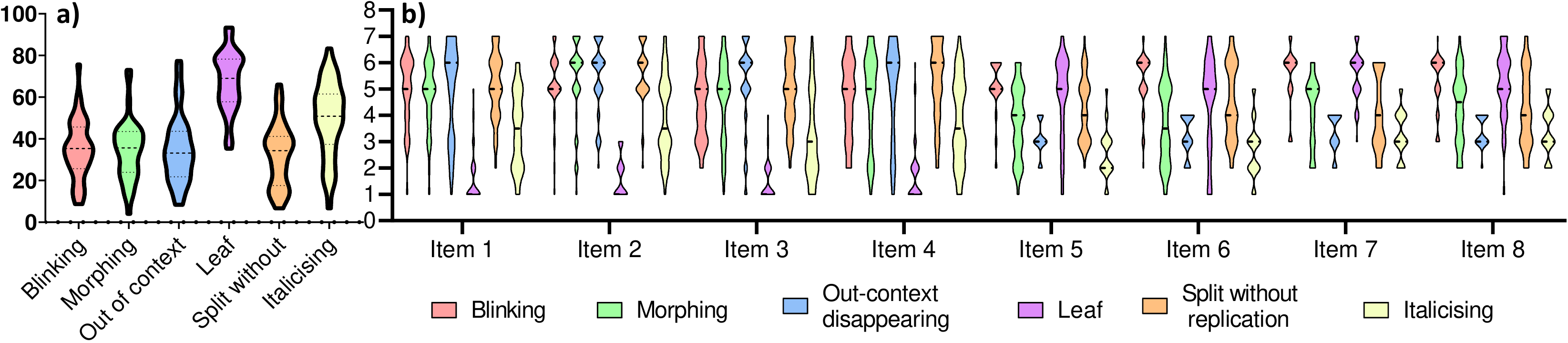}
  \vspace{-8pt}
  \caption{a) NASA-TLX weighted answers b) UEQ-S answers (Items 1 to 4 relate to PQ and items 4 to 8 relate to HQ).}
  \label{fig:UEQ-result}
  
\end{figure*}


  



Figure \ref{fig:UEQ-result} provides a visual summary of the weighted perceived cognitive load obtained from the NASA-TLX questionnaire and the answers to the UEQ-S questionnaires for each menu type. 
On the one hand, the results show that the \textit{Leaf} and \textit{Italicising} menus are considered to be more cognitively demanding than the other menu types.
O the other hand, the analysis of the answers collected from items 1 to 4 showed that the PQ of the menus can vary considerably. 
In particular, the \textit{Leaf} and \textit{Italicising} menus can be considered to be obstructive, complicated, inefficient and confusing when selecting a menu item from a prediction window. 
On the contrary, the \textit{Out context disappearing} and \textit{Blinking} menus were considered to be very helpful as regards highlighting the prediction window. This means that menus with different structures than usual and menus with different font types (\eg italics) are perceived to be worse in terms of ease of use when compared to than motion-changing or position-changing menus.
The analysis of the answers collected from items 5 to 8 suggest that almost every menu was perceived as a visually attractive menu. Only the \textit{Out of context disappearing} and \textit{Italicising} menus were perceived as unattractive. 
Note that the pragmatic quality and the perceived cognitive load are inversely correlated (\ie when pragmatic quality increases, perceived cognitive load decreases).
These findings suggest that some menus are considered to be attractive but difficult to use (\eg the \textit{Leaf} menu), or attractive and easy to use (\eg the \textit{Split without replication} menu) or unattractive and difficult to use (\eg the \textit{Italicising} menu). These different dimensions could, therefore, be useful when designing UIs that provide a good perceived UX.

After analysing the questionnaire results, we performed the correlation analysis. First, the Shapiro–Wilk test performed for the six menu types revealed a normal distribution for all the variables. We consequently applied Pearson’s correlation coefficient in order to compare the means
of the emotion-based measures with the corresponding perception-based measures. Table \ref{tab:correlation} provides a summary of the correlation analysis results. We computed the mean attraction and cognitive load for each of the menus and correlated these values with the mean PA and PCL, respectively.  
These results suggest that hypotheses $H_{n21}$ and $H_{n22}$ can be rejected, signifying that attraction and cognitive load are highly correlated with perceived attraction and perceived cognitive load, respectively. This could mean that EEG signals may effectively be used to evaluate some dimensions of user experience (\ie cognitive load and attraction).

    \begin{table}
          \caption{Hypotheses testing}
          \label{tab:hypothesis-test}
          \begin{tabular}
          {m{0.218\columnwidth}
          m{0.225\columnwidth}
          m{0.115\columnwidth}
          m{0.13\columnwidth}
          m{0.10\columnwidth}
          }
            \toprule
            Variable & Method & p-value & Cliff's $\delta$ estimate & Effect size \\
            \midrule
            Cognitive load & Kruskal-Wallis & 0.000 & 0.585 & large \\
            
            Engagement & Kruskal-Wallis & 0.000 & 0.277 & medium \\
            
            Attraction & Kruskal-Wallis & 0.000 & 0.302 &  medium\\
            
            Memorisation & Kruskal-Wallis & 0.000 &  0.667 & large\\
            
            CT & Kruskal-Wallis & 0.000 & 0.575 & large\\
            
          \bottomrule
        \end{tabular}
        \end{table}


\subsection{Discussion}


    
    



    With regard to $RQ1$,
    the results showed that there were statistically significant differences among different graphical adaptive menus. 
    Moreover, the analysis was complemented with the magnitude of their effect size. Cognitive load, memorisation and the completion time had a large effect size, while attraction and engagement had a medium effect size. These results suggest that the adaptive menus studied have not only statistically significant differences, but also practical significant differences.
    We also discovered some trends that may be useful when deciding which menus to use in software systems in order to increase the UX. The menus which are more different may be more difficult to use in a real application owing to uncertainty as to how the users will react to them, while the menus which produced a more similar reaction could predict the users' reactions. For example, and as shown in Figure \ref{fig:emotionBasedResults1}, the use of menus with different font styles, movement or structures, may lead to more uncertainty,
    and menus with colour and temporal variance may lead to a more predictable reaction.
    The results obtained were similar to those from the baseline experiment. We selected a specific context in order to study whether the differences discovered in the baseline experiment were caused by the difference in the participants’ background rather than by the differences in the menus. This experiment made it possible to confirm that the differences in UX are caused by the menu design.


    With regard to $RQ2$,
    we can conclude that emotion-based attraction and cognitive load are highly correlated with perception-based attraction and cognitive load, respectively. 
    This correlation suggests that the metrics calculated from the EEG are accurate, signifying that EEG devices could be used to obtain UX measurements faster and in a more effective manner than traditional questionnaires. 

    The emotion-based, performance-based and perception-based variables showed that the debate concerning preference \vs performance is always ongoing~\cite{Nielsen1994PrefvsPerf}. In this experiment, we were able to discover that some users prefer, or find more attractive, menus which do not increase performance. For example, menus with high ratings for both Attraction and Perceived Attraction also registered a high cognitive load when compared to other menus (\eg the \textit{Leaf} menu).
    We consider that these findings can be used by software designers who wish to apply adaptive menus to their software. When deciding which menu to use, those that registered greater differences should be avoided. We also consider that this information could be used in adaptive systems, since the bio-information on how humans react to software could improve the adaptation rules so as to improve UX through the use of different graphical menus.

    \begin{table}
          \caption{Correlation analysis results}

          \label{tab:correlation}
          \begin{tabular}
          {m{0.22\columnwidth}
          m{0.15\columnwidth}
          m{0.115\columnwidth}
          m{0.13\columnwidth}
          m{0.10\columnwidth}
          }
            \toprule
            Menu & Attraction & PA & Cognitive load & PCL \\
            \midrule

Blinking & 11.336 & 1.306 & 29.420 & 29.983 \\
Morphing & 4.225 & -0.018 & 35.321 & 41.758 \\
 Out of context 
 & -1.159 & -0.725 & 30.189 & 32.275 \\
Leaf & 3.331 & 1.020 & 35.444 & 66.933 \\
Split without 
& 0.755 & 0.281 & 30.672 & 33.758 \\
Italicising & -1.002 & -1.037 & 33.351 & 43.816 \\
       
            \specialrule{0.5pt}{3pt}{3pt}
            
            \multicolumn{2}{l}{Variables} &\multicolumn{2}{r}{Correlation} & \multicolumn{1}{r}{p-value} \\

             \multicolumn{3}{l}{Cognitive load - PCL} & \multicolumn{1}{r}{0.817} & \multicolumn{1}{r}{0.046} \\
             \multicolumn{3}{l}{Attraction - PA}  & \multicolumn{1}{r}{0.819} & \multicolumn{1}{r}{0.045} \\

          \bottomrule
        \end{tabular}
        \end{table}

\subsection{Threats to validity}

\textit{Internal threats} - The learning effect was mitigated by randomising the order in which the menus were used. The understandability of the slides used to represent the menus was assessed during the experiment, after the calibration phase, as described in Section \ref{ref:prepDA}. Finally, in order to avoid any 
source of bias, the experimental materials were evaluated by an experienced researcher in Human-Computer Interaction. 

\textit{External threats} -  The representativeness of the results could have been affected by the number of menus that were compared. We believe that the menus selected for this study can be considered as a baseline with which to obtain indications as to which properties produce better results on UX. However, we are aware that they cannot represent the whole design space of graphical adaptive menus. Replications with different and more complex graphical menus are nevertheless necessary in order to study the effect of the graphical adaptive menu variables on the results obtained.

\textit{Construct threats} - 
    The reliability of the questionnaire as regards assessing the PA was tested using  Cronbach’s alpha test.
    With the exception of the Cronbach's alpha test regarding the HQ of the \textit{Leaf} menu, all the results were higher than the threshold level (0.70)~\cite{Maxwell2002Applied}.
    PCL was measured using the NASA-TLX survey instrument, which is widely used to measure cognitive load. 

\textit{Conclusion threats} - The same data extraction procedure was systematically applied to each participant. The electrodes were placed correctly according to the guidelines~\cite{electrode1994guideline}, and the proper statistical tests were performed for the variables and their assumptions~\cite{Maxwell2002Applied}.

\section{Conclusions and Future Work}

    This paper presents an internal replication of an experiment whose objective was to analyse how 20 graphical adaptive menus impacted on the participants' UX, measured by using an EEG device. We additionally investigated the correlation between the EEG signals and the participants’ UX ratings. The results showed that there were statistically significant differences among the participants' UX (\ie cognitive load, engagement, memorisation and attraction) when interacting with the different graphical adaptive menus. Moreover, large and medium practical differences were found for all the variables. These results confirm the results obtained in the baseline experiment with a more homogeneous group of users, and the effect size shows that these results could be generalised to other contexts.
    
    The results also showed that the emotion-based measures are highly correlated with the perception-based measures, suggesting that EEG signals may be used to evaluate certain dimensions of UX (\ie attraction and cognitive load). We also found certain trends, such as the fact that menus that include different font styles, movement or structures may lead to more uncertainty with respect to their impact on UX. Moreover, menus with colour and temporal variance may lead to a more predictable user reaction. Furthermore, we found that menus that are preferred by the users do not necessarily perform well.
    These results may be useful for any software designers interested in the effect of different graphical adaptive menus on the emotions and cognitive load of end users, and for software developers when defining or selecting adaptation mechanisms for UIs.
    
    As future work, we plan to increase the knowledge regarding how users react to different UI elements by running additional EEG experiments. This will allow us to provide some recommendations on how to adapt user interfaces of software systems in order to better fit with user needs in different contexts.

\begin{acks}

This work was supported by the AKILA project (CIAICO/2021/303) funded by the GVA and the European Union through the Operational Program of the European Social Fund (ESF). D. Gaspar-Figueiredo is recipient of a Predoctoral Research staff-training program (GVA ACIF/2021/172) funded by the GVA.



\end{acks}

\bibliographystyle{ACM-Reference-Format}
\bibliography{references-file}


\end{document}